\documentclass[final,5p,times,twocolumn]{elsarticle}
\usepackage{graphicx} 
\usepackage{amsmath,amssymb,bm} 
\usepackage{siunitx}  
\usepackage{physics}  
\usepackage{tikz}     

\usepackage{lineno}
\modulolinenumbers[5]
\nolinenumbers

\journal{Journal of \LaTeX\ Templates}

\begin{document}

\begin{frontmatter}

\title{Assessment of the Imaging Performance of the CITIUS High-Resolution Detector for Heavy Charged Particles and Neutrons}

\author[icepp]{Y. Kamiya\corref{corresp}}
\ead{kamiya@icepp.s.u-tokyo.ac.jp}
\cortext[corresp]{Corresponding author}

\author[JASRI,RIKEN]{H. Nishino}
\author[RIKEN]{T. Hatsui}

\address[icepp]{Department of Physics and International Center for Elementary Particle Physics, The University of Tokyo, 7-3-1 Hongo, Bunkyo-ku, Tokyo 113-0033, Japan}
\address[JASRI]{Japan Synchrotron Radiation Research Institute (JASRI), 1-1-1 Kouto, Sayo-cho, Sayo-gun, Hyogo 679-5198, Japan}
\address[RIKEN]{RIKEN SPring-8 Center, 1-1-1 Kouto, Sayo-cho, Sayo-gun, Hyogo 679-5148, Japan}

\begin{abstract}
We report on the assessment of the imaging performance of CITIUS -- a high-speed X-ray detector developed 
for the large-scale synchrotron radiation facility SPring-8-II -- 
for heavy charged particles and neutrons.
To characterize the detector response, an irradiation experiment was performed using alpha particles 
from an $^{241}$Am source at four back-bias voltages of \SI{400}{V}, \SI{300}{V}, \SI{200}{V}, and \SI{170}{V}, 
thereby controlling the amount of charge diffusion. 
A Geant4 model of the experiment was constructed, 
and four model parameters were determined 
by template fitting to the measured signal cluster shape distributions. 
The best-fit values are: an intrinsic energy spread of 5\% for the source,
a gold fraction of 0.4 for the Au--Pd coating, 
a lateral charge diffusion spread of \SI{26.5}{\um} over a drift distance of \SI{650}{\um} at \SI{400}{V} back-bias,
and a per-pixel readout noise of \SI{10000}{e^-} in the medium-gain channel.
Using the obtained sensor model, simulations were performed for \SI{4}{\MeV} alpha particles and cold neutrons 
to evaluate the expected spatial resolution. 
In both cases, simulated CITIUS, when operated in a gain-selecting mode between high and medium gains, 
yields a substantial improvement:
at a pixel size of \SI{70}{\um} for example, the resolution improves from \SI{9.1}{\um} to \SI{1.2}{\um} for alpha particles, 
and from \SI{26}{\um} to \SI{1.9}{\um} for cold neutrons. 
These results suggest that two key features of CITIUS 
-- its gain-selecting architecture and the substantial charge sharing enabled by the long carrier drift distance -- 
extend its imaging capabilities beyond X-rays to heavy charged particles and neutrons.
\end{abstract}

\begin{keyword}
Silicon pixel detector; Alpha particle imaging; Neutron imaging; Spatial resolution; Geant4
\end{keyword}

\end{frontmatter}


\section{Introduction}

Recent advances in X-ray imaging detectors have been driven by devices composed of thick silicon sensors integrated 
with CMOS (complementary metal-oxide-semiconductor) readout circuitry. 
For imaging heavier charged particles -- such as protons and alpha particles -- as well as neutrons, 
the thick silicon sensor with long carrier drift distances leading to substantial charge sharing among neighbouring pixels, 
is expected to yield excellent spatial resolution.

CITIUS -- a high-speed X-ray detector developed for the large-scale synchrotron radiation facility SPring-8-II\cite{spring8-II} -- 
is one such X-ray imaging detector and has demonstrated best-in-class performance 
in high-resolution, high-sensitivity X-ray ptychographic coherent diffraction imaging\cite{citius}. 
In this paper, we evaluate its response and imaging performance for alpha particles and neutrons.
To quantitatively understand the detection performance for these particles, 
experimental characterization of the detector response under high-injection conditions 
-- where the transient charge density exceeds the impurity concentration in the silicon sensor -- 
is essential, as existing mobility and recombination models in this regime often lack sufficient accuracy. 

In this study, we performed an irradiation experiment using alpha particles from an $^{241}$Am source.
This allowed us to extract the back-bias dependence of charge diffusion and readout noise, 
with a view toward neutron detection via the $^{10}$B$(n,\alpha)^{7}$Li reaction. 
On the basis of these results, a detector simulation model was developed to evaluate the spatial resolution 
for alpha particles and neutrons as a function of pixel size and noise level.

The paper is organized as follows. 
Section 2 describes the alpha-particle irradiation experiment and the template-fitting procedure. 
Section 3 presents the spatial-resolution evaluation for alpha particles and neutrons using the simulation model.
Section 4 provides a summary.

\section{Determination of the Diffusion and Noise Level} 
\subsection{alpha-particle irradiation experiment} 

The CITIUS sensor consists of square pixels with a pitch of $\SI{72.6}{\micro\meter}$, 
arranged in an array of 728 pixels in the $x$ direction and 384 pixels in the $y$ direction, 
providing an active area of approximately $\SI{52.9}{\mm} \times \SI{27.9}{\mm}$.
Three different gain stages -- high-gain, medium-gain, and low-gain -- are implemented in the readout.
CITIUS adopts a gain-selecting architecture, in which appropriate gain values are chosen 
by digital logic implemented at the periphery of the CMOS circuitry. 
This approach also improves the effective noise performance for charge clusters, 
as the optimal noise level can be selected according to the charge collected at each pixel.
This is expected to provide a significant advantage when determining 
the particle incident position as the charge-weighted centroid of the cluster.

\begin{figure}[thb]
\begin{center}
\includegraphics[width=0.68\linewidth]{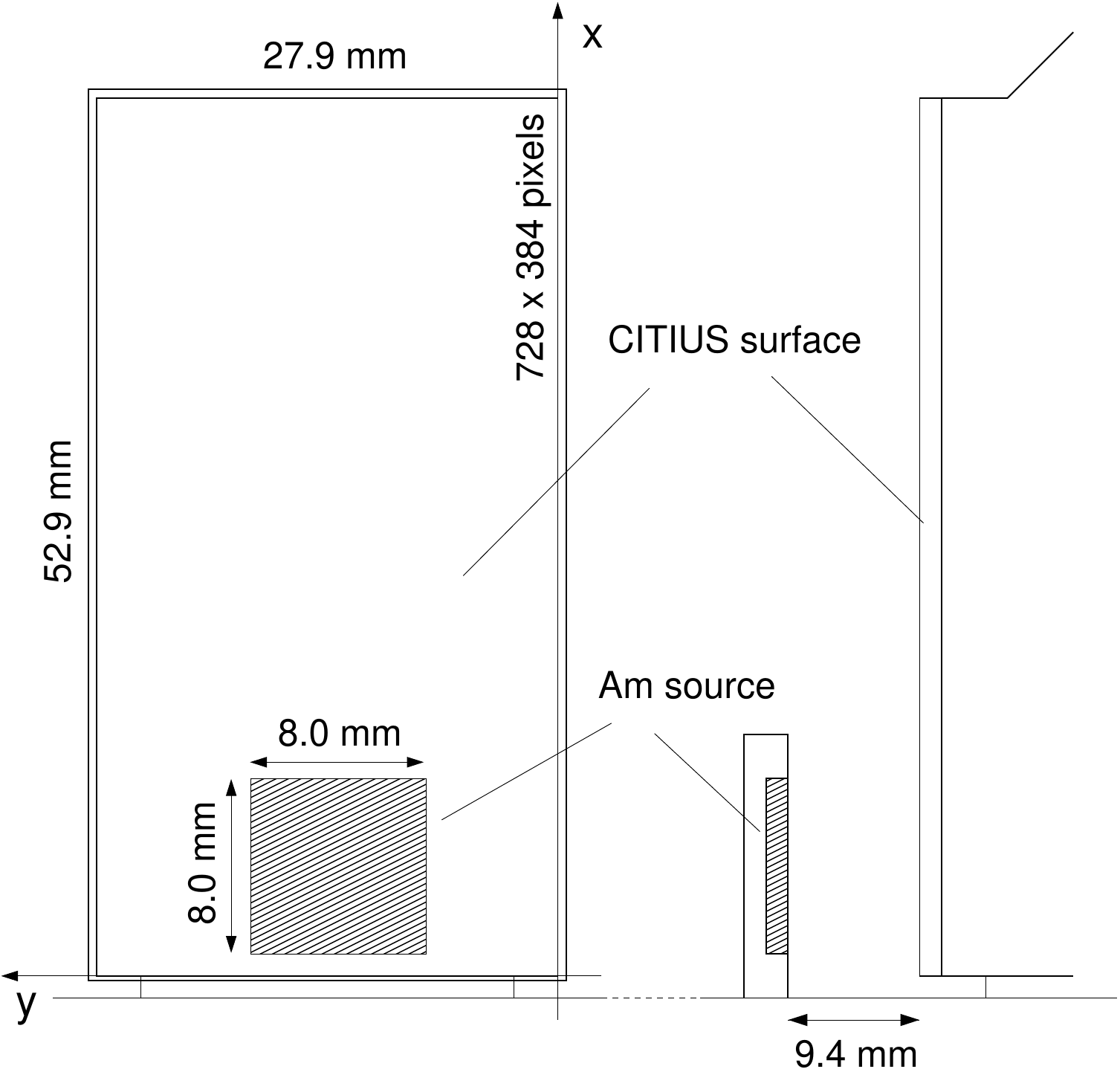}
\caption{
Experimental setup. The $^{241}$Am source has an $8 \times \SI{8}{\mm^2}$ active area and is placed \SI{9.4}{\mm} from the CITIUS sensor surface. 
Alpha particles lose energy in the Au–-Pd coating of the source, the air gap, and the dead layer on the sensor surface before reaching the active region.
}
\label{fig:setup}
\end{center}
\end{figure}

In the present experiment, measurements were performed with CITIUS operated in a mode 
where only medium-gain values were read out, hereafter referred to as the single-gain configuration. 
This mode enables characterization of charge diffusion under high-injection conditions with minimal ambiguity. 
For comparison, simulations were also conducted for a gain-selecting mode employing both high and medium gains, referred to as the multi-gain configuration.
Data were acquired at four back-bias voltages $V_b$: \SI{400}{V}, \SI{300}{V}, \SI{200}{V}, and \SI{170}{V}.
This allowed to span from above to below the full-depletion voltage of \SI{200}{V} for the \SI{650}{\um}-thick sensor.
Figure \ref{fig:setup} shows a schematic overview of the experimental setup.
The experiment was conducted in air, 
with the CITIUS sensor maintained at \SI{30}{\degreeCelsius}. 
Data were acquired at a frame rate of 17.4 kHz, with \SI{19960}{frames} collected for each back-bias voltage.
%

The $^{241}$Am radiation source, 
with the active deposit distributed over an $\SI{8}{\mm} \times \SI{8}{\mm}$ area,
was placed at a distance of \SI{9.4}{\mm} from the CITIUS surface, 
and alpha particles were directed onto the back side of the sensor.
The source is coated in a \SI{3}{\um}-thick Au--Pd alloy layer of unknown composition, 
through which the alpha particles lose part of their energy before entering the air gap between the source and the sensor.
After passing through the air layer, the alpha particles interact with the CITIUS sensor, 
which consists of a topmost insensitive dielectric layer on a 650-\SI{}{\um}-thick reverse-biased depleted silicon substrate. 
The thickness of the dielectric layer is negligible compared to the range of alpha particles. 
Grazing-incidence X-ray quantum efficiency measurements confirm that the sensitive (active) region extends very close to the dielectric/silicon interface, 
with the insensitive silicon region being less than \SI{3}{\nm} thick. 
The combined dead layer, comprising the dielectric layer and the insensitive silicon region, is therefore negligible for alpha particles.
In the present experiment, alpha particles arrive at the active region with energies below \SI{5}{\MeV}, 
corresponding to a range in silicon of less than approximately \SI{25}{\um},
meaning that most of the energy is deposited near the back surface of the sensor.
The charge generated there drifts toward the sensing nodes while diffusing 
through the \SI{650}{\um}-thick silicon layer, 
and is digitized by on-chip ADCs.
The bit depth in the operating mode used in this study was 12 bits.
When operated in high-gain mode with the ADC in coarse mode, 
the RMS noise was approximately 1 ADC count (1 DN: Digital Number), 
which is low enough for precise charge measurement at each pixel.
The same holds for the medium- and low-gain modes under the same conditions.
A detailed description of the readout architecture will be reported elsewhere.

The signal forms a cluster that spreads over multiple pixels and exhibits a shape close to a two-dimensional Gaussian with a single peak.
A $7 \times 7$ pixel window centered on the local peak pixel is used to define the cluster.
In this experiment, event clustering was performed with a threshold of 330 keV for the local peak pixel.
The number of cluster events obtained for each back-bias voltage was \SI{450093}{events} for $V_b = \SI{400}{V}$,
\SI{450695}{events} for $V_b = \SI{300}{V}$, \SI{421302}{events} for $V_b = \SI{200}{V}$, and \SI{414571}{events} for $V_b = \SI{170}{V}$.
As described in Refs. \cite{b10int4,b10int4-sw}, 
the cluster shape is characterized by the following three quantities, 
where $q_i$ and $\vec{r}_i$ denote the charge and position of the $i$th pixel, respectively:
\begin{eqnarray}
m_0 &=& \sum_{i \in \mathrm{window}} q_i ,\\
\vec{m}_1 &=& \frac{1}{m_0} \sum_{i \in \mathrm{window}} q_i \vec{r}_i ,\\
\vec{m}_2 &=& \frac{1}{m_0} \sum_{i \in \mathrm{window}} q_i (\vec{r}_i - \vec{m}_1)^{\circ 2} .
\end{eqnarray}
Here, the product of vectors on the right-hand side of Eq. (3) denotes the Hadamard (element-wise) product.

\begin{figure*}[thb]
\begin{center}
\includegraphics[width=0.98\linewidth]{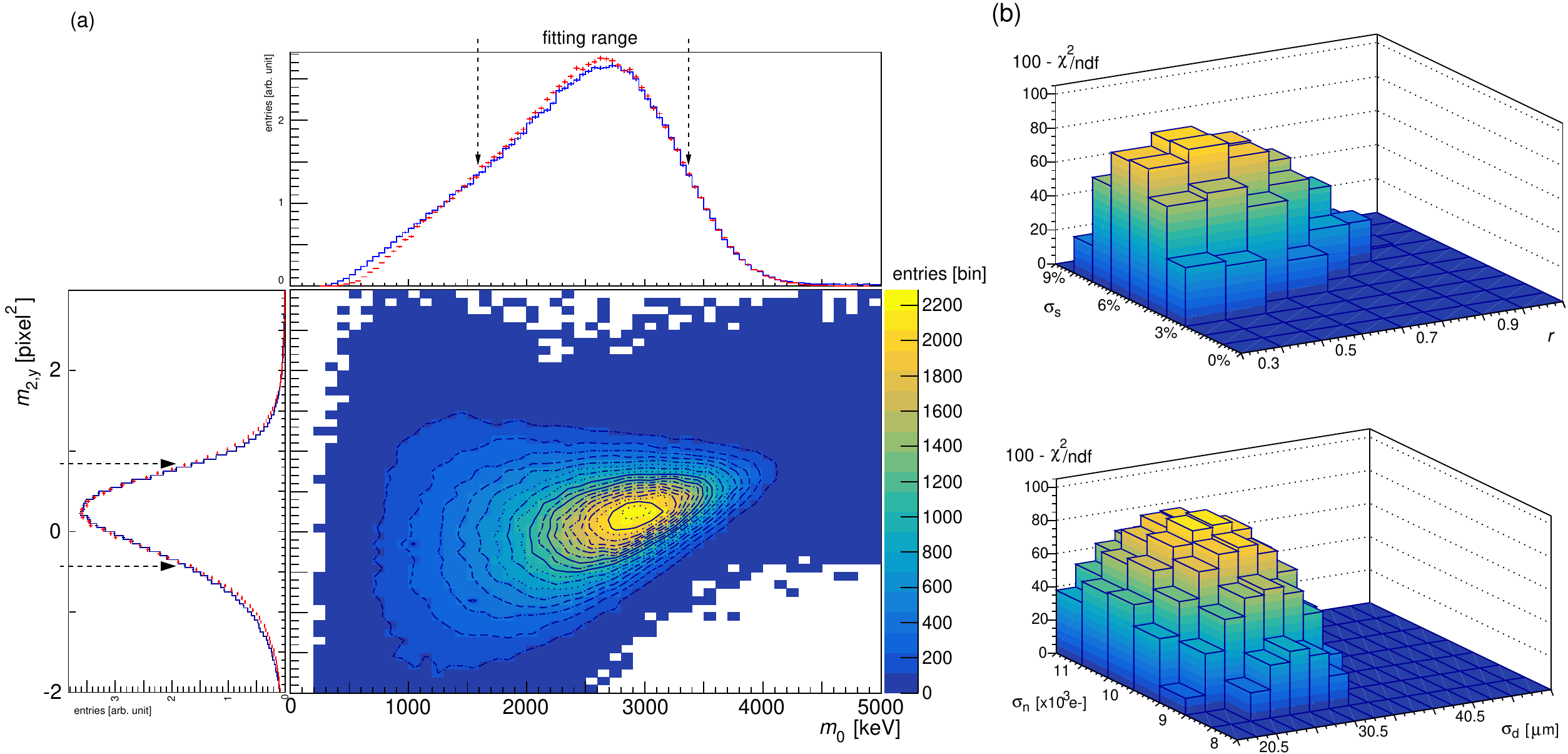}
\caption{
(a) Measured distribution of $m_{2,y}$ versus $m_0$ for $V_b = \SI{400}{V}$, 
with the projected distributions shown as histograms along the top and left axes.
Cross markers indicate the best-fit template distribution. 
Dashed arrows indicate the fitting range, defined as the bins within the half-maximum of the distribution peak.
(b) $\chi^2$ distributions for the source parameters (top) and the sensor parameters (bottom).
The source parameter $\chi^2$ is obtained by summing over all four back-bias conditions. 
The sensor parameter $\chi^2$ is shown for $V_b = \SI{400}{V}$ as a representative example.
For clarity, the vertical axis is plotted as $100 - \chi^2/\mathrm{ndf}$.
The best-fit values are $\sigma_s = 5\%$, $r = 0.4$, and, for $V_b = \SI{400}{V}$,
$\sigma_d = \SI{26.5}{\um}$ and $\sigma_n = \SI{10000}{e^-}$.
}
\label{fig:scatt}
\end{center}
\end{figure*}

Figure \ref{fig:scatt}(a) shows the measured distribution of $m_{2,y}$ versus $m_0$ for $V_b = \SI{400}{V}$,
with the projected distributions shown as histograms along the top and left axes.
Here $m_{2,y}$ denotes the $y$-component of $\vec{m}_2$.
The distribution of $m_{2,x}$, the $x$-component of $\vec{m}_2$, is similar to that of $m_{2,y}$.

%

\subsection{simulation model}

A model of the experiment was constructed using Geant4 v11.3.2 
with the QBBC physics list with the EMZ electromagnetic physics option to handle particle interactions with matter, 
with four free parameters to be determined from the experimental data.

Alpha particles from $^{241}$Am were generated isotropically from the $\SI{8}{\mm} \times \SI{8}{\mm}$ gold matrix with the following energies and branching ratios: 
\SI{5.486}{\MeV} (86.0\%), \SI{5.443}{\MeV} (12.7\%), and \SI{5.389}{\MeV} (1.3\%)\cite{IN-1261}.
To account for the non-uniformity of the radioactive source deposition and potential systematic effects beyond the scope of the present model,
an intrinsic energy spread $\sigma_s$ was applied as a common Gaussian broadening to all spectral lines. 
This is the first model parameter.
The \SI{3}{\um}-thick Au--Pd alloy coating of the source has an unknown composition; writing the gold fraction as $r$, 
the value of $r$ is determined from the experimental data. 
This is the second model parameter. 
These two parameters define the source model.

Within the active region of the sensor,
each electron-hole pair is to be created by \SI{3.65}{\eV} of deposited energy.
The total number of generated e-h pairs was sampled from a Gaussian distribution with a Fano factor of 0.1.
The \SI{650}{\um}-thick active layer was discretized into $10^4$ sublayers, each of thickness \SI{65}{\nm}.
Charge diffusion was modeled as a two-dimensional Gaussian 
with a spread $\sigma_{d}$, defined as the lateral broadening 
after drifting \SI{650}{\um} in the depth direction.
This is the third model parameter.
The drift length for each charge carrier was calculated from its layer position within the discretized sensor depth.
The resulting charge depositions were then projected onto the sensing nodes by integrating the two-dimensional Gaussians, 
whose widths are proportional to the square root of the drift length.
The charge at each sensing node was then sampled from a Poisson distribution 
whose mean is given by the integrated charge, while conserving the total number of generated e-h pairs.
Finally, the readout noise of each pixel was modeled as a Gaussian 
with mean zero and standard deviation $\sigma_n$.
This is the fourth model parameter and these two parameters define the sensor model.

\subsection{template fitting}
The following parameter ranges were scanned to simulate the measured signal:
\begin{eqnarray}
0 \% < &\sigma_s& < 9 \% \quad \quad \quad ~~ \text{in steps of 1 \%,} \\
0.3 < &r& < 1.0 \quad \quad \quad ~~ \text{in steps of 0.1,} \\
\SI{18.5}{\um} < &\sigma_d& < \SI{48.5}{\um} \quad ~~~ \text{in steps of \SI{2.0}{\um},} \\
\SI{8000}{e^-} < &\sigma_n& < \SI{11000}{e^-} \quad \text{in steps of \SI{500}{e^-}.}
\end{eqnarray}
For each parameter set, the distributions of $m_0$ and $m_{2,y}$
were constructed and compared with the experimental data to determine the best-fit parameters.

Among the source model parameters, $\sigma_s$ was consistently determined 
to be 5\% across all back-bias conditions.
The best-fit value of $r$ fell in the range of 0.3 to 0.5;
summing the $\chi^2$ distributions over all four back-bias voltages 
yielded $r = 0.4$ as the most probable value.
The upper histogram of Figure \ref{fig:scatt}(b) shows the $\chi^2/\mathrm{ndf}$ distribution for the source parameters,
here, $\chi^2$ is defined as
\begin{equation}
\chi^2 \equiv \sum_k \frac{(n_k - n_{k,template})^2}{\sigma_k^2},
\end{equation}
where $\sigma_k$ is the statistical uncertainty of the $k$-th bin,
and $n_k$ is the number of entries in that bin.
$n_{k,template}$ is the expected number of entries in the $k$-th bin 
from the template simulation for a given set of model parameters.
The summation was performed over the bins within the half-maximum of the distribution peak.

\begin{figure}[thb]
\begin{center}
\includegraphics[width=0.98\linewidth]{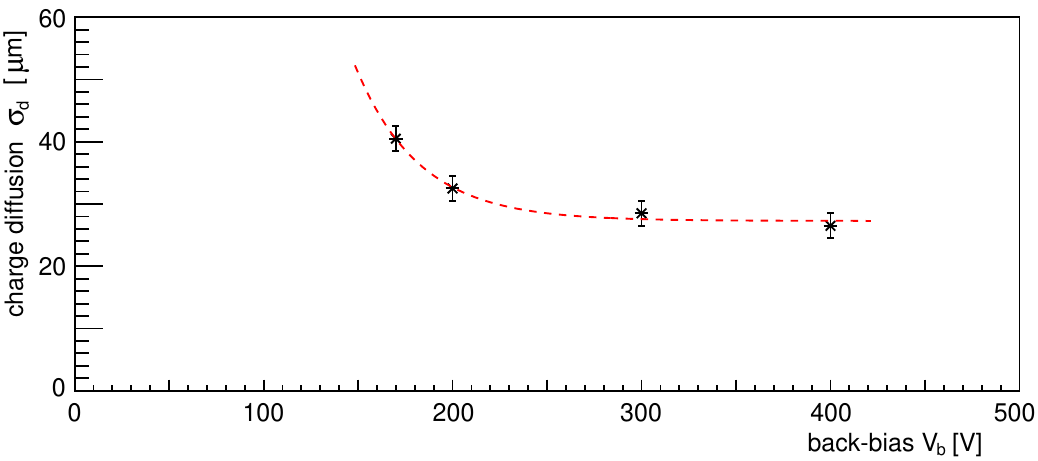}
\caption{
Back-bias dependence of the charge diffusion spread $\sigma_d$.
Sufficient charge sharing is observed across all conditions, 
suggesting that CITIUS is well suited also for alpha particle and neutron imaging.
The full depletion voltage was $V_b = \SI{200}{V}$.
}
\label{fig:diffSigma}
\end{center}
\end{figure}

The sensor parameters $\sigma_d$ and $\sigma_n$ 
were evaluated individually for each back-bias condition,
with the source parameters fixed at $\sigma_s = 5\%$ and $ r = 0.4$.
The lower histogram of Figure \ref{fig:scatt}(b) shows the $\chi^2/\mathrm{ndf}$ distribution at a back-bias of \SI{400}{V},
and the best-fit distributions are indicated by the cross markers in Figures \ref{fig:scatt}(a) top and Figures \ref{fig:scatt}(a) left.
The charge diffusion in the CITIUS sensor over a drift distance of \SI{650}{\um} 
was determined to be $\sigma_d = 26.5$, $28.5$, $32.5$, and \SI{40.5}{\um} 
for $V_b = 400$, $300$, $200$, and $\SI{170}{V}$, respectively.
The noise level was determined to be $\sigma_n = \SI{10000}{e^-}$, showed no significant variation with $V_b$.
Figure \ref{fig:diffSigma} shows the back-bias dependence of charge diffusion $\sigma_d$.
%

\section{Evaluation of the Spatial Resolution}

The spatial resolution is defined as the standard deviation obtained 
by fitting the Line Spread Function (LSF) along the $y$ direction with a Gaussian. 
To also quantify deviations from a Gaussian profile, the RMS of the LSF is used as a complementary measure. 
When the readout noise is relatively large compared to the signal, 
the LSF is well approximated by a Gaussian, and the two measures agree. 
In contrast, when the LSF has a flat-top profile, 
or when alpha particles are incident at a specific zenith angle in a low-noise environment, 
the Point Spread Function (PSF) becomes ring-shaped, and the LSF can exhibit multiple peaks. 
Such cases manifest as a discrepancy between the two measures.

\subsection{for alpha particles}

\begin{figure}[thb]
\begin{center}
\includegraphics[width=0.98\linewidth]{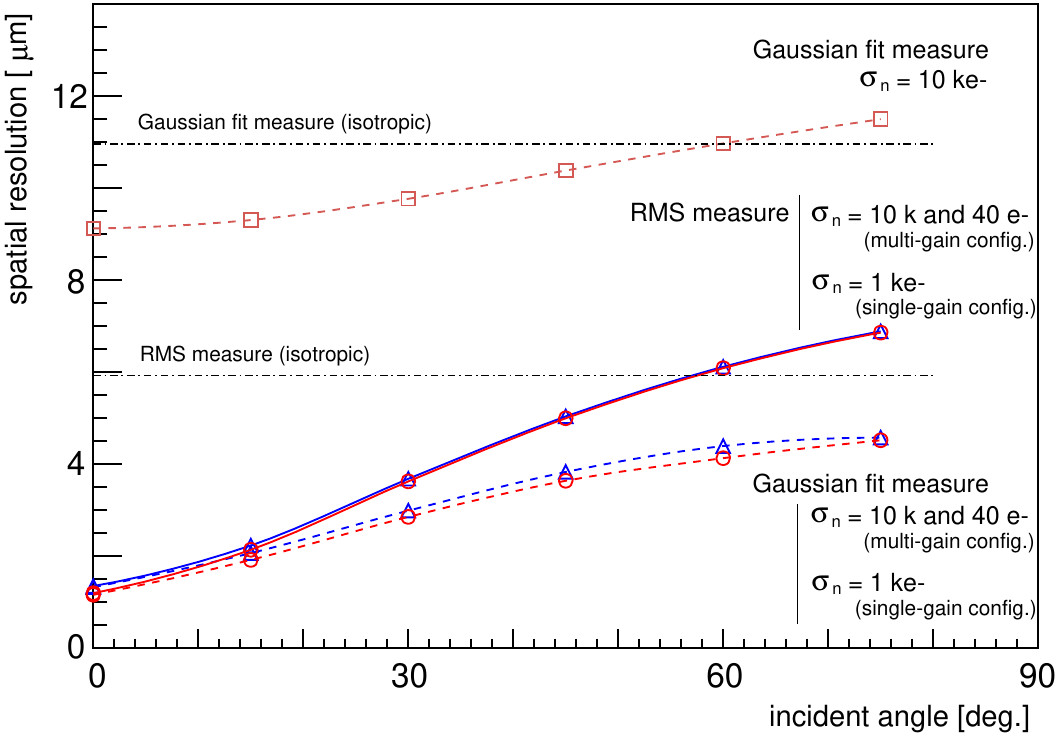}
\caption{
Spatial resolution as a function of the incident angle for alpha particles.
Open Squares, triangles, and circles represent 
the single-gain configuration with $\sigma_n = \SI{10}{ke^-}$,
the single-gain configuration with $\sigma_n = \SI{1}{ke^-}$,
and the multi-gain configuration with $\sigma_n = \SI{10}{ke^-}$ and $\SI{40}{e^-}$, respectively.
For each, solid and dashed lines represent the RMS measure and the Gaussian fit measure, respectively.
Dash-dotted lines indicate the results for isotropic incidence.
At large incident angles, a discrepancy between the two measures arises due to the ring-shaped PSF.
}
\label{fig:alphaAngle}
\end{center}
\end{figure}

We first examine the spatial resolution as a function of the incident angle.
In the simulation, monoenergetic alpha particles of \SI{4}{\MeV} are generated 
at the back surface of the sensor at a fixed incident angle,
with a pixel size of $\SI{70}{\um}$ and a back-bias of $V_b = \SI{400}{V}$ ($\sigma_d = \SI{26.5}{\um}$).
The results are shown in Figure \ref{fig:alphaAngle}.
%

In the single-gain configuration ($\sigma_n = \SI{10}{ke^-}$, open squares),
the LSF is well approximated by a Gaussian, and no significant difference is observed 
between the Gaussian fit measure and the RMS measure. 
As the incident angle decreases toward normal incidence, the spatial resolution improves, 
owing to the reduced displacement due to the particle range in the silicon.
The spatial resolution for isotropic incidence is indicated by the dash-dotted line.

The case with a noise level reduced by a factor of ten is shown by open triangles, 
where the solid and dashed lines represent the RMS measure and the Gaussian fit measure, respectively. 
Both measures give results consistent between the multi-gain configuration ($\sigma_n = \SI{10}{ke^-}$ and $\SI{40}{e^-}$)
and the single-gain configuration with $\sigma_n = \SI{1}{ke^-}$,
where, in the multi-gain configuration, 
the high-gain channel ($\sigma_n = \SI{40}{e^-}$) is used 
when the charge at the pixel node is below $\SI{65}{ke^-}$, 
and the medium-gain channel ($\sigma_n = \SI{10}{ke^-}$) is used for charges between $\SI{65}{ke^-}$ and $\SI{6.2}{Me^-}$.
At large incident angles, a significant discrepancy between the two measures arises due to the ring-shaped PSF mentioned earlier. 
As the incident angle approaches normal incidence, the ring-shaped distribution dissolves and the two measures converge.

\begin{figure}[thb]
\begin{center}
\includegraphics[width=0.98\linewidth]{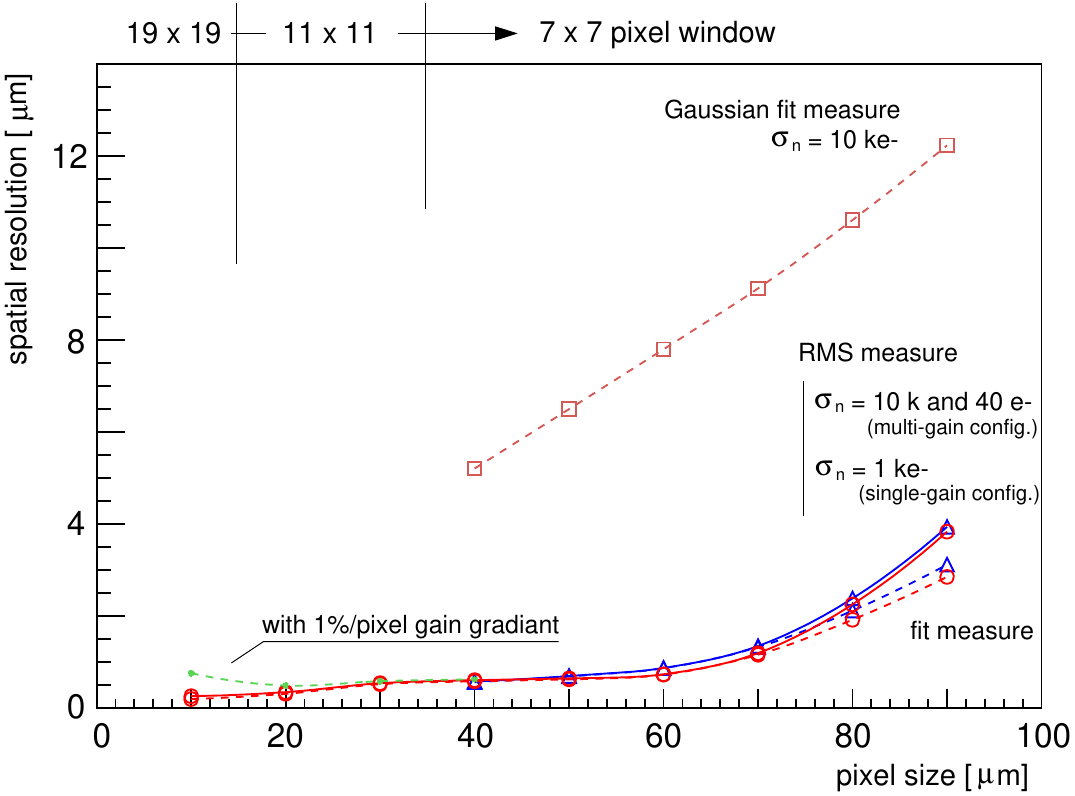}
\caption{
Spatial resolution as a function of the pixel size for alpha particles.
Open Squares, triangles, and circles represent 
the single-gain configuration with $\sigma_n = \SI{10}{ke^-}$,
the single-gain configuration with $\sigma_n = \SI{1}{ke^-}$,
and the multi-gain configuration with $\sigma_n = \SI{10}{ke^-}$ and $\SI{40}{e^-}$, respectively.
For each, solid and dashed lines represent the RMS measure and the Gaussian fit measure, respectively.
Small filled circles represent the effect of systematic pixel-to-pixel non-uniformity, 
which may arise from gain or charge transport variations,
as an example, with a random gain gradient $\delta / \mathrm{pixel}$ 
in one direction, uniformly distributed in the range $0.5\% < \delta < 1.5\%$.
The pixel window size used for each pixel size is indicated along the top of the figure.
}
\label{fig:alphaLSF}
\end{center}
\end{figure}

Figure \ref{fig:alphaLSF} shows the spatial resolution as a function of pixel size for normal incidence. 
As in the previous figure, the Gaussian fit measure and the RMS measure agree 
in the single-gain configuration ($\sigma_n = \SI{10}{ke^-}$, open squares). 
The single-gain configuration with $\sigma_n = \SI{1}{ke^-}$ (open triangles) also shows 
behavior consistent with that of the multi-gain configuration.
For smaller pixel sizes, a larger pixel window is used to fully capture the tails of the cluster: 
$19 \times 19$ pixels for a pixel size of \SI{10}{\um}, and $11 \times 11$ pixels for \SI{20}{\um} and \SI{30}{\um}.
At larger pixel sizes, a discrepancy between the Gaussian fit measure and the RMS measure becomes apparent, 
reflecting the onset of non-smooth charge sharing distributions due to the discrete pixel structure.

In the charge-weighted centroid method used here, 
the spatial resolution can surpass the intrinsic single-pixel resolution -- pixel size /$\sqrt{12}$ --
thanks to charge sharing across multiple pixels. 
Since the achievable precision depends critically on accurate charge measurements in the low-signal pixels away from the local maximum, 
the gain-selecting architecture of CITIUS, which enables the use of a low-noise readout channel for these pixels in the tails of the cluster, 
is considered to be the key factor behind the substantial improvement in spatial resolution.
This also suggests that further improvement may be achievable by introducing a dynamic cluster windowing algorithm 
that optimizes the window shape on an event-by-event basis, rather than the fixed-size window employed in the present analysis.

The present simulation assumes an ideal two-dimensional Gaussian distribution for charge drift and diffusion; 
a detailed investigation of the limitations of this model is beyond the scope of the present paper 
and would require careful comparison with experimental data. 
As one example of such effects, the impact of systematic pixel-to-pixel non-uniformity 
-- which may arise from gain or charge transport variations -- 
was investigated by introducing, for each cluster, a random gain gradient $\delta / \mathrm{pixel}$ in one direction, 
with a magnitude uniformly distributed in the range $0.5\% < \delta < 1.5\%$.
The results are shown by small filled circles, suggesting that this effect contributes at the sub-micrometer level.

\subsection{for neutron particles}

To make the sensor sensitive to neutrons, a boron layer is deposited on the back surface of the sensor 
to convert neutrons to charged particles via the $^{10}$B$(n,\alpha)^{7}$Li reaction:
\begin{eqnarray*}
n + {}^{10}{\rm B} \!\!\!
&\rightarrow& \!\!\! \alpha_{(\SI{1.47}{\MeV})} + {}^{7}{\rm Li}_{(\SI{0.84}{\MeV})} + \gamma_{(\SI{0.48}{\MeV})} \quad \text{\small{(93.9\%)}},\\
&\rightarrow& \!\!\! \alpha_{(\SI{1.78}{\MeV})} + {}^{7}{\rm Li}_{(\SI{1.01}{MeV})} \quad \text{\small{(6.1\%)}}.
\end{eqnarray*}
In the dominant branch, the gamma-ray is emitted upon de-excitation of the recoiling $^{7}$Li nucleus.
Since its energy is small compared to the mass of the nucleus, 
the recoil imparted to the $^{7}$Li is negligible, and the two heavy charged particles are emitted in nearly opposite directions. 
It is therefore expected that one of the two particles enters the CITIUS sensor.

In the present simulation, the boron layer thickness is set to \SI{200}{\nm}
to facilitate comparison with previous studies\cite{b10int4,b10int4-sw},
where the corresponding boron layer was deposited by Ar RF sputtering.
The incident beam is modeled with a kinetic energy of 2.53 meV (velocity 695 m/s), 
a representative energy in the cold neutron range widely used in condensed matter research.
Since the absorption length in enriched $^{10}$B is of the order of several \SI{}{\um},
neutron capture reactions are expected to occur nearly uniformly throughout the depth of the boron layer. 
Accordingly, alpha particles or $^{7}$Li recoils are generated uniformly within the $^{10}$B layer and emitted isotropically.

\begin{figure}[thb]
\begin{center}
\includegraphics[width=0.88\linewidth]{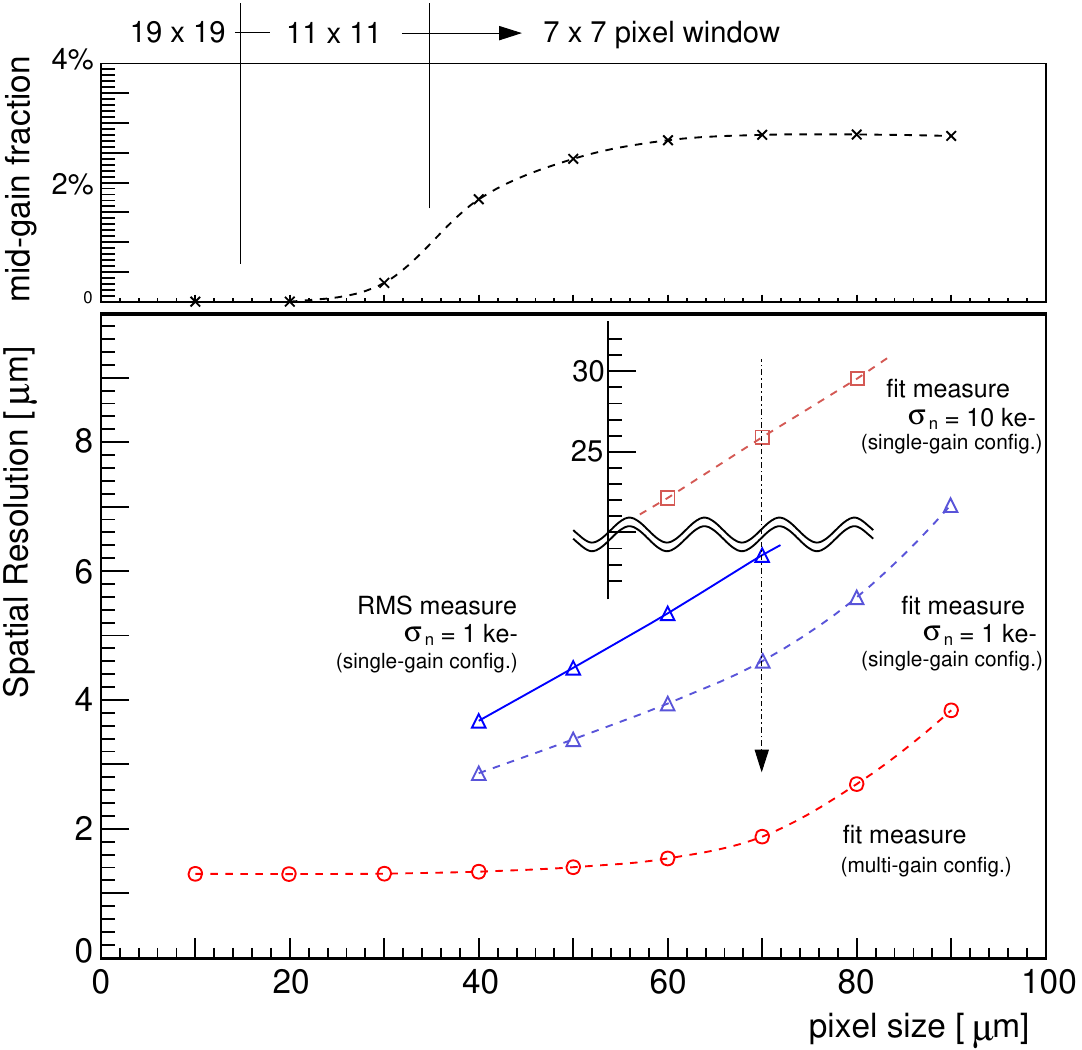}
\caption{
Spatial resolution as a function of the pixel size for neutrons.
Open Squares, triangles, and circles represent 
the single-gain configuration with $\sigma_n = \SI{10}{ke^-}$,
the single-gain configuration with $\sigma_n = \SI{1}{ke^-}$,
and the multi-gain configuration with $\sigma_n = \SI{10}{ke^-}$ and $\SI{40}{e^-}$, respectively.
For each, solid and dashed lines represent the RMS measure and the Gaussian fit measure, respectively.
Significant improvement by the gain-selecting architecture is observed.
The upper panel shows the mean fraction of pixels within the cluster window 
for which the medium-gain channel signal is used in the multi-gain configuration. 
The pixel window size used for each pixel size is indicated along the top of the figure.
}
\label{fig:b10LSF}
\end{center}
\end{figure}

Figure \ref{fig:b10LSF} shows the pixel size dependence of the spatial resolution for neutrons.
As in the previous figure, the single-gain configuration ($\sigma_n = \SI{10}{ke^-}$, open squares),
the single-gain configuration with reduced noise ($\sigma_n = \SI{1}{ke^-}$, open triangles),
and the multi-gain configuration ($\sigma_n = \SI{10}{ke^-}$ and $\SI{40}{e^-}$, open circles) are shown, 
with solid and dashed lines representing the RMS measure and the Gaussian fit measure, respectively.
As before, a larger pixel window is used for smaller pixel sizes.
%

\begin{figure}[thb]
\begin{center}
\includegraphics[width=0.88\linewidth]{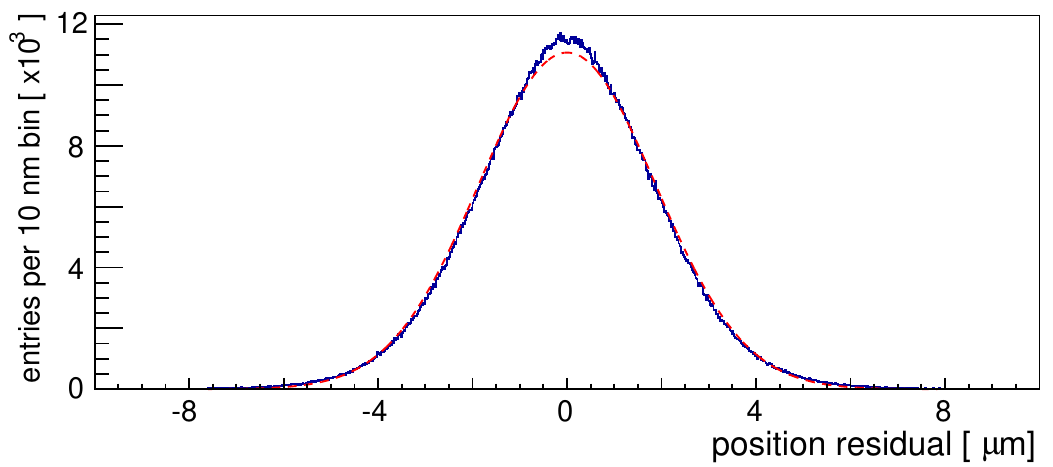}
\caption{
The solid histogram shows the Line Spread Function 
along the $y$ direction for the multi-gain configuration at a pixel size of \SI{70}{\um}.
The vertical axis shows the number of simulated entries per 10 nm bin.
The dashed line shows the corresponding Gaussian fit, with $\sigma = \SI{1.9}{\um}$.}
\label{fig:b10fit}
\end{center}
\end{figure}

In both the single-gain ($\sigma_n = \SI{10}{ke^-}$),
and multi-gain configurations,
the discrepancy between the two measures remains below 1\% 
and is therefore not shown in the figure.
In the single-gain configuration with $\sigma_n = \SI{1}{ke^-}$, 
visible difference between the two measures is observed, 
which arises from the double-Gaussian shape of the LSF in neutron detection, 
reflecting contributions from both the alpha particle and the $^{7}$Li recoil.
Figure \ref{fig:b10fit} shows a representative example 
of the LSF and the corresponding Gaussian fit 
for the multi-gain configuration at a pixel size of $\SI{70}{\um}$.
%

A significant improvement in spatial resolution with the multi-gain configuration is also evident for neutron detection.
As in the case of alpha particles, the precise charge measurement in the low-signal pixels 
at the tails of the cluster is considered to be a key factor in the improvement of the spatial resolution. 
The upper panel of Figure \ref{fig:b10LSF} shows the mean fraction of pixels within the cluster window 
for which the medium-gain channel signal is used in the multi-gain configuration. 
For pixel sizes of \SI{40}{\um} and above, only 2--3 pixels near the cluster center 
carry sufficient charge to be read out via the medium-gain channel. 
For pixel sizes of \SI{30}{\um} and below, sufficient charge sharing ensures 
that the charge per pixel falls within the dynamic range of the high-gain channel.

\section{Summary}
To characterize the response of the CITIUS sensor to heavy charged particles, 
an irradiation experiment was performed using alpha particles from an $^{241}$Am source 
at four back-bias voltages $V_b$ of \SI{400}{V}, \SI{300}{V}, \SI{200}{V}, and \SI{170}{V}. 
A model of the experiment was constructed using Geant4, and four model parameters 
to the measured signal cluster shape distributions. 
The best-fit values are: an intrinsic energy spread of $\sigma_s = 5\%$ for the source, a gold fraction of $r = 0.4$ for the Au-Pd coating,
a lateral charge diffusion spread of $\sigma_d = \SI{26.5}{\um}$ over a drift distance of \SI{650}{\um} at $V_b = \SI{400}{V}$, 
and a per-pixel readout noise of $\sigma_n = \SI{10000}{e^-}$ in the medium-gain channel. 
Even at the highest back-bias of \SI{400}{V}, sufficient charge sharing is expected across the sensor.

Using the obtained sensor model, simulations were performed for \SI{4}{\MeV} alpha particles and cold neutrons. 
In both cases, the gain-selecting architecture yields a substantial improvement in spatial resolution: 
at a pixel size of \SI{70}{\um}, the resolution improves from \SI{9.1}{\um} to \SI{1.2}{\um} for alpha particles, 
and from \SI{26}{\um} to \SI{1.9}{\um} for cold neutrons. 
These results strongly suggest that CITIUS, originally developed as a high-speed X-ray imaging detector, 
is also well suited for imaging applications involving heavy charged particles and neutrons.
Demonstration experiments are currently in preparation.

\section*{Acknowledgements}
This work was supported by JSPS KAKENHI Grant Number 23H00106 
and by the TIA Kakehashi framework in 2025 and 2024.
The data that support the findings of this study are available from the corresponding author upon reasonable request.

\section*{Declaration of generative AI and AI-assisted technologies in the writing process}
During the preparation of this work the authors used the Claude -- Sonnet 4.6, and ChatGPT 5.3 in order to improve language. 
After using this tool/service, the authors reviewed and edited the content 
as needed and take full responsibility for the content of the publication.

\bibliography{HSTD14}

@preamble{"\providecommand{\noopsort}[1]{}" #
   "\providecommand{\singleletter}[1]{#1}%"}

@article{citius, 
year = {2023}, 
title = {{High‐resolution and high‐sensitivity X‐ray ptychographic coherent diffraction imaging using the CITIUS detector}}, 
author = {Takahashi, Yukio and Abe, Masaki and Uematsu, Hideshi and Takazawa, Shuntaro and Sasaki, Yuhei and Ishiguro, Nozomu and Ozaki, Kyosuke and Honjo, Yoshiaki and Nishino, Haruki and Kobayashi, Kazuo and Hiraki, Toshiyuki Nishiyama and Joti, Yasumasa and Hatsui, Takaki}, 
journal = {Journal of Synchrotron Radiation}, 
issn = {1600-5775}, 
doi = {10.1107/s1600577523004897}, 
pmid = {37526992}, 
pmcid = {PMC10481278}, 
pages = {989--994}, 
number = {5}, 
volume = {30}
}

@article{spring8-II, 
year = {2024}, 
title = {{Green upgrading of SPring‐8 to produce stable, ultrabrilliant hard X‐ray beams}}, 
author = {Tanaka, Hitoshi and Watanabe, Takahiro and Abe, Toshinori and Azumi, Noriyoshi and Aoki, Tsuyoshi and Dewa, Hideki and Fujita, Takahiro and Fukami, Kenji and Fukui, Toru and Hara, Toru and Hiraiwa, Toshihiko and Imamura, Kei and Inagaki, Takahiro and Iwai, Eito and Kagamihata, Akihiro and Kawase, Morihiro and Kida, Yuichiro and Kondo, Chikara and Maesaka, Hirokazu and Magome, Tamotsu and Masaki, Mitsuhiro and Masuda, Takemasa and Matsubara, Shinichi and Matsui, Sakuo and Ohshima, Takashi and Oishi, Masaya and Seike, Takamitsu and Shoji, Masazumi and Soutome, Kouichi and Sugimoto, Takashi and Suzuki, Shinji and Tajima, Minori and Takano, Shiro and Tamura, Kazuhiro and Tanaka, Takashi and Taniuchi, Tsutomu and Taniuchi, Yukiko and Togawa, Kazuaki and Tomai, Takato and Ueda, Yosuke and Yamaguchi, Hiroshi and Yabashi, Makina and Ishikawa, Tetsuya}, 
journal = {Journal of Synchrotron Radiation}, 
issn = {1600-5775}, 
doi = {10.1107/s1600577524008348}, 
pages = {1420--1437}, 
number = {6}, 
volume = {31}, 
}

@article{b10int4, 
year = {2020}, 
title = {{Development of a neutron imaging sensor using INTPIX4-SOI pixelated silicon devices}}, 
author = {Kamiya, Y. and Miyoshi, T. and Iwase, H. and Inada, T. and Mizushima, A. and Mita, Y. and Shimazoe, K. and Tanaka, H. and Kurachi, I. and Arai, Y.}, 
journal = {Nuclear Instruments and Methods in Physics Research Section A: Accelerators, Spectrometers, Detectors and Associated Equipment}, 
issn = {0168-9002}, 
doi = {10.1016/j.nima.2020.164400}, 
pages = {164400}, 
volume = {979}, 
}

@article{b10int4-sw, 
year = {2024}, 
title = {{Development of two-dimensional neutron imager with a sandwich configuration}}, 
author = {Kamiya, Y. and Nishimura, R. and Mitsui, S. and Wang, Z. and Morris, C.L. and Makela, M. and Clayton, S.M. and Baldwin, J.K. and Ito, T.M. and Akamatsu, S. and Iwase, H. and Arai, Y. and Murata, J. and Asai, S.}, 
journal = {Nuclear Instruments and Methods in Physics Research Section A: Accelerators, Spectrometers, Detectors and Associated Equipment}, 
issn = {0168-9002}, 
doi = {10.1016/j.nima.2024.169390}, 
pages = {169390}, 
volume = {1064}, 
}

@article{IN-1261, 
year = {1970}, 
title = {{Catalogue of Semiconductor Alpha-Particle Spectra}}, 
author = {R. N. Chanda and R. A. Deal}, 
journal = {IN--1261 (1970)}, 
}

\end{document}